\documentclass[global,twocolumn]{svjour}

\usepackage{graphicx}
\usepackage{color}

\journalname{Applied Physics B}
\begin{document}

\title{Birefringence of interferential mirrors at normal incidence}
\subtitle{Experimental and computational study}
\author{F. Bielsa\inst{1}, A. Dupays\inst{2,3}, M. Fouch\'e\inst{2,3}, R. Battesti\inst{1},
C. Robilliard\inst{2,3} \and C. Rizzo\inst{2,3}
}                     
\offprints{carlo.rizzo@irsamc.ups-tlse.fr}

\institute{Laboratoire National des Champs Magn\'etiques Intenses
(UPR 3228, CNRS-INSA-UJF-UPS), F-31400 Toulouse Cedex, France \and
Universit\'e de Toulouse, UPS, Laboratoire Collisions Agr\'egats
R\'eactivit\'e, IRSAMC, F-31062 Toulouse, France \and CNRS, UMR
5589, F-31062 Toulouse, France}

\date{Received: date / Revised version: date}

\maketitle
\begin{abstract}
In this paper we present a review of the existing data on
interferential mirror birefringence. We also report new measurements
of two sets of mirrors that confirm that mirror phase retardation
per reflection decreases when mirror reflectivity increases. We
finally developed a computational code to calculate the expected
phase retardation per reflection as a function of the total number
of layers constituting the mirror. Different cases have been studied
and we have compared computational results with the trend of the
experimental data. Our study indicates that the origin of the mirror
intrinsic birefringence can be ascribed to the reflecting layers
close to the substrate.
\end{abstract}
\section{Introduction}
\label{intro}

In the last decades, high reflectivity interferential mirrors have
been widely used in optical cavities to measure small light
polarization variations induced by the propagation in a weakly
anisotropic medium such as in parity violation experiments
\cite{Bouchiat19821984,Bennett1999} or in vacuum magnetic
birefringence experiments \cite{Cameron1993,Chen2007,Zavattini2008}.
Mirrors themselves are birefringent and this is manifestly a problem
for such a kind of applications because they induce a phase
retardation \cite{NotePhaseRetardation} which superimposes to the
signal to be measured. This birefringence is due to off-normal
incidence and/or to intrinsic birefringence of the mirror coatings.
In the case of Fabry-Perot cavities the incidence on the mirrors is
normal. In this paper we focus on this type of device, thus on
birefringence due to the mirror coatings.

Interferential mirrors are composed of a stack of slabs deposited on
a substrate. One slab corresponds to a low-index layer and a
high-index layer with an optical thickness $\lambda/4$ for each
layer, where $\lambda$ is the light wavelength for which the mirror
reflectivity is optimized. While non birefringent stratified media
are discussed in textbooks \cite{BornWolf}, and films with a non
trivial dielectric tensor have been treated in literature (see e.g.
\cite{Sprokel1984}), as far as we know, the origin of the mirror
birefringence is unknown, and a detailed study of the problem does
not exist. In ref. \cite{Mansuripur1997} computational results are
given in the hypothesis that the birefringence is due to only one
layer, in particular the uppermost. The author notices that the
phase retardation effect diminishes as he moves the only phase
retardation layer down the stack. In ref. \cite{Hall2000}
measurements of the mirror phase retardation as a function of time
and of laser power in the Fabry-Perot cavity have been performed.
The authors suggest that mirror birefringence may be photoinduced,
at least partly.

In this paper we present a review of the existing data on
interferential mirror phase retardation. We show that the data
indicate that the phase retardation per reflection decreases when
the mirror reflectivity becomes better and better i.e. when the
total number of layers increases. We also report new measurements of
two sets of mirrors that confirm this trend. We finally developed a
computational code to calculate the expected phase retardation per
reflection as a function of the total number of layers. Different
cases have been studied going from a fixed birefringence for each
layer to a random birefringence for each layer. We finally compare
computational results with the trend of the experimental data. Our
study indicates that the origin of the mirror intrinsic
birefringence can be ascribed to the reflecting layers close to the
substrate.

\section{Experimental study}
\label{Exp}

Birefringence of interferential mirrors have been measured and
reported by several authors
\cite{Bouchiat1982,Carusotto1989,Micossi1993,Jacob1995,Ni1996,Wood1996,Moriwaki1997,Brandi1997,Hall2000}.
The phase retardation per reflection ranges between a few $10^{-7}$
rad to $10^{-3}$ for values of $(1-R)$ going from a few $10^{-5}$ to
$10^{-2}$, where $R$ is the mirror reflectivity. All the
measurements have been conducted using an optical cavity except one
\cite{Micossi1993}, where the ellipticity acquired after a single
reflection was directly measured. Optical cavities are usually
absolutely necessary to accumulate the effect and thus to allow to
measure very small phase retardations. Whereas a multipass cavity
has been used in refs. \cite{Bouchiat1982,Carusotto1989}, a
Fabry-Perot cavity is used in refs.
\cite{Jacob1995,Ni1996,Wood1996,Moriwaki1997,Brandi1997}. In the
following section, the published data are presented in details, in
chronological order. These studies were always motivated by
measurements of small phase retardation such as parity violation
experiments \cite{Bouchiat19821984,Bennett1999} or vacuum magnetic
birefringence experiment \cite{Cameron1993}.

\subsection{Review of published data}
\label{RevExp}

The first study of intrinsic phase retardation of interferential
mirrors dates from 1982 \cite{Bouchiat1982}. Measurements have been
conducted using a multipass cavity made of two spherical mirrors
between which the light beam bounces many times forwards and
backwards under quasinormal incidence. Intrinsic phase retardation
is therefore superimposed to the off-normal incidence phase
retardation but this has also been evaluated by the authors. The
light beam does not hit the same point of the mirror after a round
trip. Thus the measurement of phase retardation per reflection gives
a value averaged on the mirror surface. The mirrors have been
manufactured by Spectra-Physics, Inc. (Mountain View, CA, USA), and
their reflectivity $R$ is 0.998 at $\lambda=$540\,nm. Intrinsic
phase retardation typically varies between 2 and 4\,$\times 10^{-4}$
rad per reflection. Among the 19 mirrors analyzed, two exceptions
with phase retardation less than $10^{-6}$ rad per reflection have
been found. The authors called this a ``happy accident".

A few years later a new study was again performed using a multipass
cavity \cite{Carusotto1989}. A set of 5\,mirrors manufactured by
MTO, Palaiseau, France, has been analyzed. The authors did not give
explicitly the reflectivity of the mirrors, but they have reported
that at $\lambda=514.5$\,nm and after about 250\,reflections the
light intensity is reduced to $1/e$. We can deduce that
$(1-R)=0.004$. From their measurements, intrinsic phase retardation
varies between $3.0\times10^{-5}$ and $2.2\times10^{-4}$ rad per
reflection.

The next study was performed in 1993 \cite{Micossi1993} using
multipass cavities. Only one mirror has been analyzed but this time
the phase retardation has been measured directly after only one
reflection. The mirror had a reflectivity of 0.9983 at 633\,nm. It
was coated by the Laboratory of Laser Energetics of the University
of Rochester. The authors were able to measure the intrinsic phase
retardation and the phase retardation axis direction of the mirror
in different points of the surface. They could therefore draw a map
of the intrinsic phase retardation showing a clear rotational
pattern. The intrinsic phase retardation per reflection ranged
between 3 to 6.2 $\times10^{-4}$ rad, while the axis direction
ranged between 9 and -13 degrees. To test that the origin of such an
anisotropy was not due to the substrate, the authors have measured
the phase retardation when the light was reflected on the
backsurface of the mirror. They obtained a result compatible with
zero within the experimental error.

In 1995 the first measurement using a Fabry-Perot cavity was
reported \cite{Jacob1995}. In this type of interferometer the
incidence on the mirrors is strictly normal, and off-normal phase
retardation vanishes. The mirror reflectivity can be inferred by the
cavity finesse $F=6600$ given by the authors at $\lambda = 633$\,nm:
$R=1-\pi/F=0.999524$. The reported values of phase retardation per
reflection are $1.0 \times 10^{-6}$ and $4.4 \times 10^{-6}$ rad.
Besides, their study allows to conclude that the birefringence is
not due to the mirror mounts.

In 1996, a new intrinsic phase retardation of a mirror is reported
\cite{Ni1996}. The Fabry-Perot cavity finesse was 300 at $\lambda =
633$\,nm, and we can therefore infer that $ R = 0.9895$. The
measured phase retardation per reflection is
$1.2\times10^{-3}$\,rad.

For the next value reported in ref. \cite{Wood1996}, a Fabry-Perot
was again used. The mirrors have been manufactured by Research
Electro-Optics Inc., Boulder, Colorado, USA. The Fabry-Perot cavity
finesse was $125600$ at $\lambda = 540$ nm, and the inferred
reflectivity is $R = 0.999975$. The value of the phase retardation
per reflection is given for only one mirror and corresponds to
$3\times10^{-6}$\,rad.

In 1997 two works have been published in the same journal issue
\cite{Moriwaki1997,Brandi1997} concerning mirror intrinsic phase
retardation. In ref. \cite{Moriwaki1997} two mirrors constituting a
Fabry-Perot cavity have been characterized. The average value of the
reported reflectivity was $R = 0.9988$ at $\lambda = 633$ nm. The
measured phase retardation per reflection was
$4.2\times10^{-4}$\,rad and $1.04\times10^{-3}$\,rad. In ref.
\cite{Brandi1997}, the reflectivity was $R = 0.999969$ at
$\lambda=1064$\,nm and they have been manufactured by Research
Electro-Optics Inc., Boulder, Colorado, USA. The measured value for
three mirrors over four was between $3.7$ and
$12\times10^{-7}$\,rad, while the last mirror was a ``happy
accident" with a phase retardation per reflection smaller than
$10^{-7}$\,rad.

Finally in 2000, a new measurement is reported \cite{Hall2000}.
Measurements have been done on a Fabry-Perot cavity, looking at
frequency shift of the resonance line of the cavity due to mirror
phase retardation. The Fabry-Perot cavity finesse was about 40 000
at $\lambda = 633$ nm, corresponding to $R = 0.999923$, and the
phase retardation per reflection $1.8\times10^{-6}$ rad. The authors
have also showed that the measured phase retardation could be
changed by several percents by appropriately injecting more power in
the cavity. Phase retardation relaxed down to the average value
several seconds after the perturbation.

In table \ref{tab:1} we summarize the existing data on mirror
intrinsic phase retardation per reflection. We give the reference
number, the value of the reflectivity $R$, the measured value of the
phase retardation per reflection $\delta_{\mathrm{M}}$, the number
of characterized mirrors $N_{\mathrm{mirrors}}$ , and finally the
light wavelength $\lambda$ for which the mirror reflectivity was
optimized. We give the minimum and the maximum value for
$\delta_{\mathrm{M}}$ when several mirrors have been analyzed in the
same reference. In the case of ref. \cite{Micossi1993}, where a
single mirror has been studied but in several points of its surface,
we give the dispersion of the reported values.

\begin{table}[h]
\caption{Review of published data.}
\label{tab:1}       
\begin{tabular}{lllll}
\hline\noalign{\smallskip}
ref. & $R$ & $\delta_{\mathrm{M}}$ (rad) & $N_{\mathrm{mirrors}}$ & $\lambda$ (nm) \\
\noalign{\smallskip}\hline\noalign{\smallskip}
\cite{Bouchiat1982} &0.998& $(2-4)\times10^{-4}$ & 17 & 540\\
       && $< 10^{-6}$& 2 & 540\\
\cite{Carusotto1989} &0.996& $(3-22)\times10^{-5}$ & 5 & 514\\
\cite{Micossi1993} &0.9983& $(3-6.2)\times10^{-4}$ & 1 & 633\\
\cite{Jacob1995} &0.999524& $(1-4.4)\times10^{-6}$ & 2 & 633\\
\cite{Ni1996} &0.9895& $1.2\times10^{-3}$ & 1 & 633\\
\cite{Wood1996} &0.999975& $3\times10^{-6}$ & 1 & 540\\
\cite{Moriwaki1997} &0.9988& $(4.2-10.4)\times10^{-4}$ & 2 & 633\\
\cite{Brandi1997} &0.999969& $(7.4-24)\times10^{-7}$ & 3 & 1064\\
                  && $< 10^{-7}$ & 1 & 1064\\
\cite{Hall2000} &0.999923& $1.8\times10^{-6}$ & 1 & 633\\
\noalign{\smallskip}\hline
\end{tabular}
\vspace*{0cm}  
\end{table}

\subsection{Our new measurements}
\label{MesBMV}

In this paragraph we report new measurements of two different sets
of mirror performed in the framework of the BMV experiment
\cite{Battesti2008} which goal is to measure vacuum magnetic
birefringence. As in the previous attempts to measure such a weak
quantity \cite{Cameron1993,Chen2007,Zavattini2008}, mirror intrinsic
phase retardation is a source of noise limiting the sensitivity of
the apparatus. Moreover, since signal detection in the BMV
experiment corresponds to a homodyne technique, the ellipticity
$\Gamma$ induced on the linearly polarized laser beam by the
Fabry-Perot cavity overall phase retardation is used as a D.C.
carrier. To reach a shot noise limited sensitivity, one needs
$\Gamma$ to be as small as possible \cite{Battesti2008}, implying
that the phase retardation axis of the two mirrors constituting the
cavity have to be aligned.

To measure the mirror intrinsic phase retardation, our experimental
method is based on the ones described in details in ref.
\cite{Jacob1995,Brandi1997}. More details on our experimental setup
can be found in ref. \cite{Battesti2008}. Briefly, 30 mW of a
linearly polarized Nd:YAG ($\lambda= 1064$ nm) laser beam is
injected into a Fabry-Perot cavity. This laser is locked to the
cavity resonance frequency using the Pound-Drever-Hall method
\cite{Drever1983}. The beam transmitted by the cavity is then
analyzed by a polarizer crossed at maximum extinction and collected
by a low noise photodiode with a noise equivalent power of
$0.25\,$pW$/\sqrt{\mathrm{Hz}}$. Polarizer extinction is $(4 \pm
2)\times 10^{-7}$ which is always much lower than the ellipticity we
measure.

\begin{figure}
\begin{center}
\resizebox{1\columnwidth}{!}{
\includegraphics{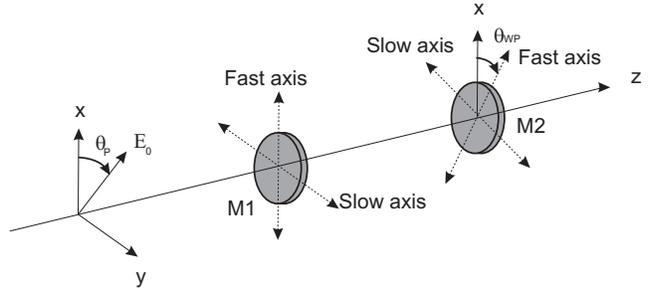}
} \caption{Principle of the experiment: a linearly polarized laser
beam is injected into a Fabry-Perot cavity (mirrors M1 and M2). The
polarization is then analyzed outside of the cavity.} \label{scheme}
\end{center}
\end{figure}

As shown on Fig.\,\ref{scheme}, both mirrors are schematized as two
ideal waveplates with phase retardation $\delta_1$ and $\delta_2$.
Thus the phase retardation per reflection of each mirror we want to
measure corresponds to $2\delta_1$ and $2\delta_2$. For the sake of
simplicity the angle indicating the direction of the phase
retardation axis of the first mirror is taken as zero. The angle
between the phase retardation axis of the two mirrors is
$\theta_\mathrm{WP}$. For $\delta_1, \delta_2 \ll 1$, combination of
both waveplates gives a single waveplate of phase retardation
\cite{Brandi1997}:
\begin{equation}\label{deltaEQ}
    \delta_\mathrm{EQ}=
    \sqrt{(\left(\delta_1-\delta_2\right)^2+4\delta_1\delta_2\cos^2\theta_\mathrm{WP}},
\end{equation}
and with a fast axis at an angle with respect to the $x$ axis given
by:
\begin{equation}\label{tetaEQ}
    \cos2\theta_\mathrm{EQ}=\frac{\frac{\delta_1}{\delta_2}+\cos2\theta_\mathrm{WP}}{\sqrt{\left(\frac{\delta_1}{\delta_2}-1\right)^2+4\frac{\delta_1}{\delta_2}cos^2\theta_\mathrm{WP}}}.
\end{equation}
The Fabry-Perot cavity corresponds to a waveplate with a phase
retardation $\delta$ related to $\delta_\mathrm{EQ}$ as follows :
\begin{equation}\label{delta}
    \delta=\frac{2F}{\pi}\delta_\mathrm{EQ},
\end{equation}
where $F$ is the cavity finesse. Finally, the intensity transmitted
by the analyzer over the incident intensity is equal to the square
of the ellipticity $\psi$ induced by the cavity mirrors. This
ellipticity is given by \cite{Brandi1997}:
\begin{equation}\label{Ext}
    \psi^2=\frac{\delta^2}{4}\sin^2(2(\theta_\mathrm{P}-\theta_\mathrm{EQ})),
\end{equation}
with $\theta_\mathrm{P}$ the angle indicating the direction of the
light polarization with respect to the $x$ axis. Thus, by measuring
the intensity transmitted by the analyzer as a function of
$\theta_\mathrm{WP}$ and for different value of $\theta_\mathrm{P}$,
we are able to calculate the phase retardation of both mirrors.

Two different sets of mirrors have been tested. The first one is
constituted by two one inch diameter spherical mirrors, 6\,m radius
of curvature, BK7 substrate, manufactured by Laseroptik GmbH,
Garbsen (Germany). The reflectivity at $\lambda = 1064$\,nm is
0.999396 corresponding to a cavity finesse of 5200 and the
transmission of the cavity is about $20\,\%$. The second set of
mirrors is constituted by three one inch diameter spherical mirrors,
8\,m radius of curvature, BK7 substrate, manufactured by Layertec
GmbH, Mellingen (Germany). The reflectivity at $\lambda = 1064$\,nm
is 0.999972 corresponding to a cavity finesse of about 110000.
According to the manufacturer, mirror losses are lower than
$100\,$ppm and the overall measured transmission of the cavity is
about $3\,\%$.

\begin{figure}
\begin{center}
\resizebox{1\columnwidth}{!}{
\includegraphics{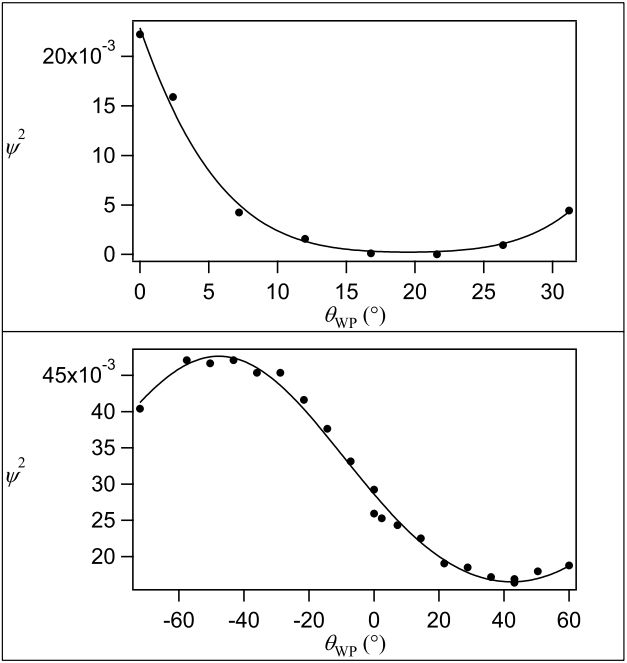}
} \caption{Experimental values of the square of the ellipticity
$\psi$ as a function of the angle between the phase retardation axis
of the cavity mirrors (see table \ref{tab:ellipticity}). Data are
fitted using Eq.\,(\ref{Ext}). Upper curve: the mirrors reflectivity
is 0.999396. Lower curve: the mirrors reflectivity is 0.999972.}
\label{fig:ellipticity}
\end{center}
\end{figure}

The square of the ellipticity $\psi$ induced by the cavity as a
function of the angle between the phase retardation axis of the two
mirrors is plotted in Fig.\,\ref{fig:ellipticity}. Experimental
values are fitted using Eq.\,(\ref{Ext}). The deduced mirror
intrinsic phase retardation per reflection is presented in Table
\ref{tab:ellipticity} for each mirror.

\begin{table}[h]
\caption{Mirror intrinsic phase retardation.}
\label{tab:ellipticity}       
\begin{tabular}{llll}
\hline\noalign{\smallskip}
$R$ & $\delta_\mathrm{M}$ (rad)& No. & $\lambda$ (nm) \\
\noalign{\smallskip}\hline\noalign{\smallskip}
0.999396 & $(5.8\pm0.4)\times10^{-4}$ & 1 & 1064\\
          & $(3.4\pm0.4)\times10^{-4}$ & 2 & \\
0.999972 & $(9.8\pm0.4)\times10^{-6}$ & 1 & 1064\\
          & $(2.6\pm0.4)\times10^{-6}$ & 2 & \\
          & $(1\pm0.4)\times10^{-6}$ & 3 & \\
\noalign{\smallskip}\hline
\end{tabular}
\vspace*{0cm}  
\end{table}

\subsection{Summary}
\label{ExpSum}

\begin{figure}
\begin{center}
\resizebox{1\columnwidth}{!}{
\includegraphics{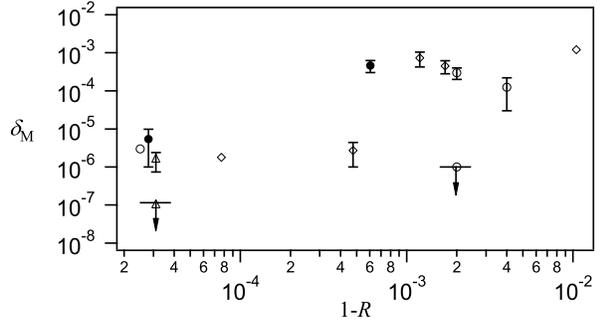}
} \caption{Summary of all the published data and the data obtained
in this work with mirror intrinsic phase retardation
$\delta_\mathrm{M}$ versus $(1-R)$. The symbols represent the
wavelength for which the mirror reflectivity was optimized ($\circ$
: 540 nm, $\diamond$ : 633 nm, {\tiny $\triangle$} : 1064 nm,
$\bullet$ : our work). Errors bars correspond to the minimum and the
maximum value when several mirrors have been analyzed. Arrows
represent mirrors for which the phase retardation was smaller than
the apparatus sensitivity. The trend of the whole points shows that
the intrinsic phase retardation decreases by 3 orders of magnitude
as $(1-R)$ decreases by almost 3 orders of magnitude.}
\label{deltavsR}
\end{center}
\end{figure}

All the published data together with the data obtained in this work
are plotted as a function of $(1-R)$ on Fig.\,\ref{deltavsR}. When
only one mirror has been tested, the corresponding point has no
error bars. When different mirrors have been measured the data point
have error bars. These error bars do not represent the measurement
error for one mirror (typically 10\%) but the dispersion of the
measured value for the whole set of mirrors. Arrows represent
mirrors for which the phase retardation was smaller than the
apparatus sensitivity (see table \ref{tab:1} and
\ref{tab:ellipticity}). Dots represent the new measurements reported
in this work at $\lambda$ = 1064 nm.

Published data plotted on Fig.\,\ref{deltavsR} clearly show that the
higher the reflectivity i.e. the lower the value of $(1-R)$, the
lower the phase retardation per reflection. More precisely, the
intrinsic phase retardation decreases by 3 orders of magnitude as
$(1-R)$ decreases by almost 3 orders of magnitude. Our new
measurements perfectly confirm this trend.

\section{Computational study}
\label{Comp}

The understanding of the origin of the experimental data trend is
crucial if one wants to control the manufacture to obtain
birefringence free interferential mirrors. We have therefore
developed a computer code that can simulate the behavior of an
interferential mirror made by an arbitrary number of layers each one
with its own arbitrary phase retardation and arbitrary retardation
axis. Our goal was to find a configuration of layers, as simple as
possible, that could reproduce the experimental trend and give a
first indication to experimentalists to test in further studies.

\subsection{Interferential mirrors}

\begin{figure}
\begin{center}
\resizebox{1\columnwidth}{!}{
\includegraphics{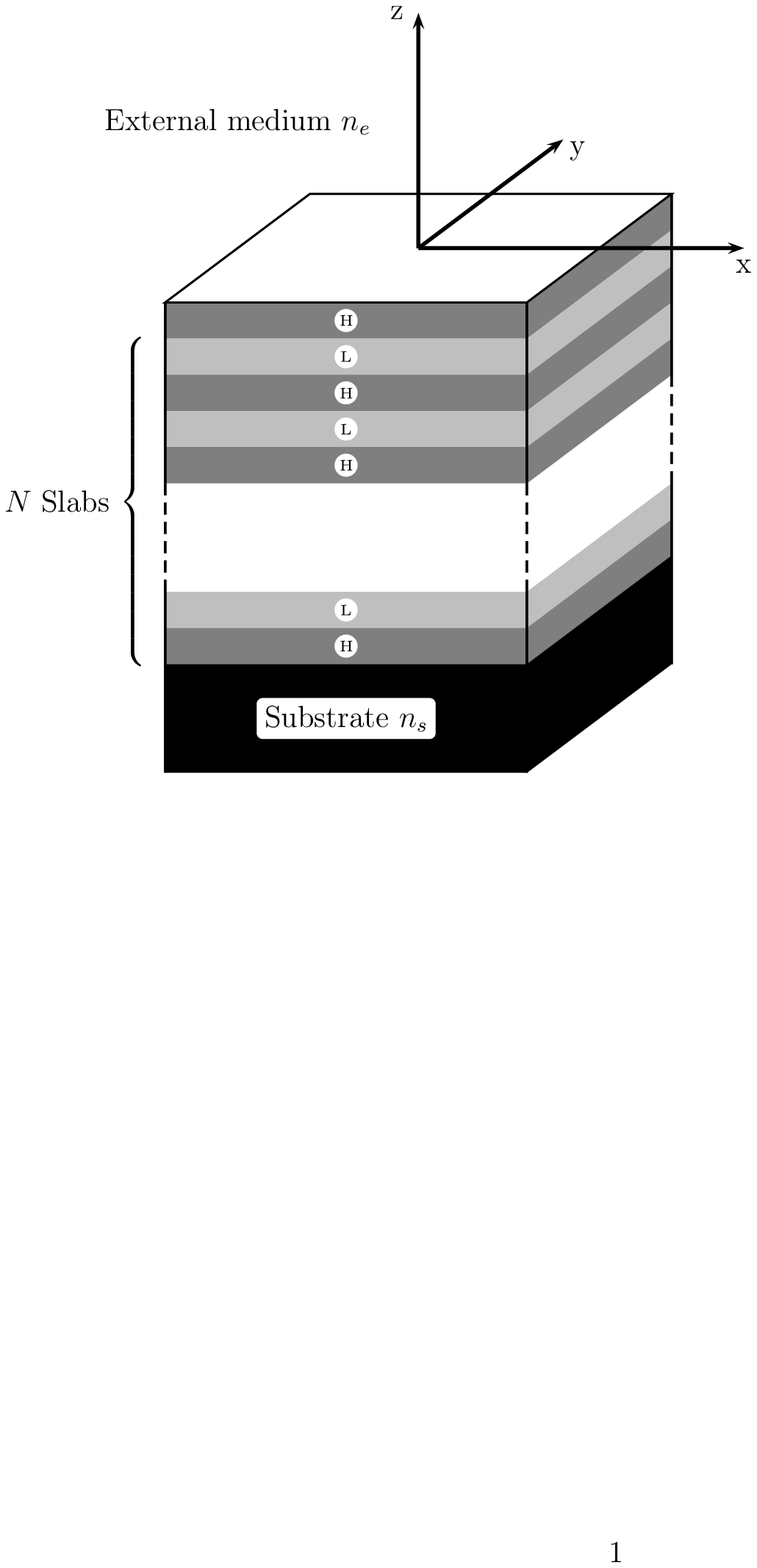}
} \caption{Interferential mirror. It consists of an ``odd stack" of
slabs deposited on a substrate.} \label{MultiLayer}
\end{center}
\end{figure}

Interferential mirrors are made by a stack of slabs of an optical
thickness of $\lambda/2$ as shown on Fig.\,\ref{MultiLayer}, where
$\lambda$ is the light wavelength for which the mirror reflectivity
is optimized. Each slab is composed by a low-index layer
$n_\mathrm{L}$ and a high-index layer $n_\mathrm{H}$. Each layer has
an optical thickness of $\lambda/4$. Typically, $n_\mathrm{L}$ is
around 1.5 and $n_\mathrm{H}$ is higher than 2.0. The substrate is
usually fused silica or Zerodur, and a $\lambda/2$ coating of
Si0$_2$ protects the reflecting surface of the mirror. Obviously,
construction details are not shared publicly by manufacturers (see
e.g. the paragraph on mirror manufacture in ref.
\cite{Stedman1997}).

In the case of what is called an ``odd stack" i.e. $N$ slabs of a
high-index layer and a low-index layer plus one high-index layer
($2N+1$ layers), the mirror reflectivity $R$ can be written as
\cite{BornWolf}:
\begin{equation}\label{RvsN}
    R = \left[\frac{1-\left(\frac{n_\mathrm{H}}{n_\mathrm{s}}\right)^2\left(\frac{n_\mathrm{H}}{n_\mathrm{L}}\right)^{2N}}{1+\left(\frac{n_\mathrm{H}}{n_\mathrm{s}}\right)^2\left(\frac{n_\mathrm{H}}{n_\mathrm{L}}\right)^{2N}}\right]^2
\end{equation}
where $n_\mathrm{s}$ is the index of refraction of the substrate.
Typically to obtain a reflectivity $R \simeq 0.999999$ one needs
about 20 pairs of quarter-wavelength layers of materials such as
SiO2 and either TiO$_2$ or TaO$_5$, while 10 pairs are sufficient to
obtain $R \simeq 0.999$.

\subsection{Methods}

The model multilayer we used for our calculations consists of a
stack of slabs placed between two semi-infinite media of refractive
indices $n_\mathrm{e}$ (the external medium) and $n_\mathrm{s}$ (the
substrate). The coordinate system used to reference the multilayer
axes is shown in Fig.\,\ref{MultiLayer}.

\label{CompMet}

\begin{figure}[h]
\begin{center}
  \includegraphics[width=6cm]{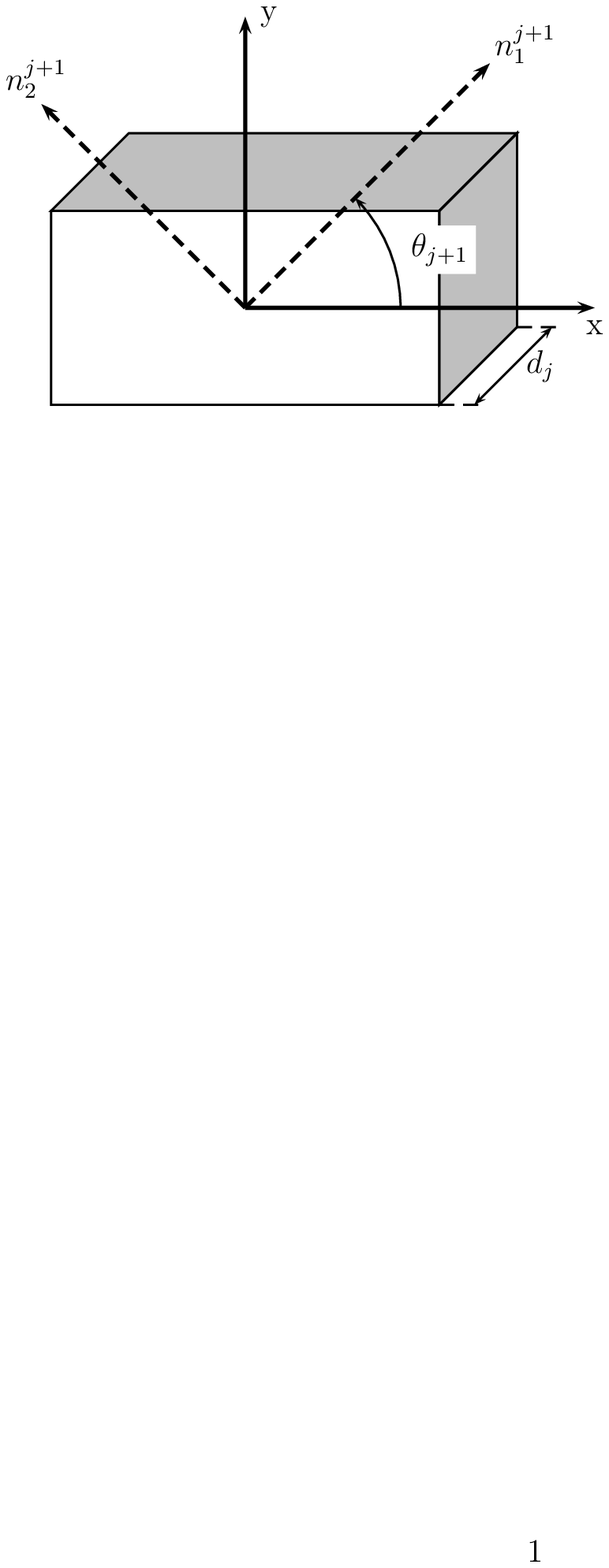}\\
  \caption{Angle between the principal axis of the birefringent medium
and the reference frame.}\label{BLayer}
\end{center}
\end{figure}

Each birefringent layer is uniaxial. For the $j$th layer extending
from $z=z_j$ to $z=z_{j+1}$ we denote by $\theta_{j+1}$ the angle
between the principal axis of the birefringent medium and the
reference frame and by $d_j=z_{j+1}-z_j$ its thickness (see
Fig.\,\ref{BLayer}).

In the reference frame, the dielectric tensor of this layer is then
given by
\begin{equation}
\epsilon^{j+1}=R^{-1}(\theta_{j+1}) \left(\begin{array}{lr}
\epsilon_1^{j+1} & 0 \\
0 & \epsilon_2^{j+1}
\end{array}\right)
R(\theta_{j+1})
\end{equation}
where $R(\theta)$ is the standard rotation matrix :
\begin{equation}
\label{rotationmatrix} R(\theta)=\left(
\begin{array}{lr}
\cos{\theta} & \sin{\theta} \\
-\sin{\theta} & \cos{\theta}
\end{array}\right).
\end{equation}
For a low-index layer, we have
\begin{equation}
\left\{
\begin{array}{l}
n_1^{j+1}=\sqrt{\epsilon_1^{j+1}/\epsilon_0}=n_\mathrm{L}+\delta n_\mathrm{L} \nonumber \\
n_2^{j+1}=\sqrt{\epsilon_2^{j+1}/\epsilon_0}=n_\mathrm{L} \nonumber
\end{array} \right.
\end{equation}
and for a high-index layer
\begin{equation}
\left\{
\begin{array}{l}
n_1^{j+1}=\sqrt{\epsilon_1^{j+1}/\epsilon_0}=n_\mathrm{H}+\delta n_\mathrm{H} \nonumber \\
n_2^{j+1}=\sqrt{\epsilon_2^{j+1}/\epsilon_0}=n_\mathrm{H} \nonumber
\end{array} \right.
\end{equation}
where $\epsilon_0$ is the vacuum permeability and $n_\mathrm{L},
n_\mathrm{H}$ stand for refractive indices of similar but
no-birefringent layers with an optical thickness of $\lambda/4$, so
that $n_{1,2}^{j+1}d_j=\lambda/4$.

Let us now consider a transverse electric polarized plane
monochromatic wave normally incident upon this model mirror. The
solution of the Maxwell's equations for the electric field can be
expressed as a superposition of the forward and backward propagating
waves along each reference direction $x$ and $y$. In the external
medium, we have
\begin{eqnarray}
\label{Electricfield}
E_{x}=&A&_{\mathrm{e},x}^+\exp\{i(k_{\mathrm{e},x}z-\omega t)\} \nonumber \\
       &+&A_{\mathrm{e},x}^-\exp\{i(-k_{\mathrm{e},x}z-\omega t)\}
\end{eqnarray}
for the $x$ component and
\begin{eqnarray}
\label{Electricfield}
E_{y}=&A&_{e,y}^+\exp\{i(k_{\mathrm{e},y}z-\omega t)\} \nonumber \\
       &+&A_{\mathrm{e},y}^-\exp\{i(-k_{\mathrm{e},y}z-\omega t)\}
\end{eqnarray}
for the $y$ component, where $\omega = 2\pi/\lambda$ and
\begin{equation}
k_{\mathrm{e},x}=k_{\mathrm{e},y}=\frac{\omega}{c}n_\mathrm{e}
\end{equation}
with $c$ the light velocity in vacuum. In the same way, the electric
field in the substrate is written as
\begin{eqnarray}
\label{Electricfield}
E_{x}=&A&_{\mathrm{s},x}^+\exp\{i(k_{\mathrm{s},x}(z-z_{2N+1})-\omega t)\} \nonumber \\
       &+&A_{\mathrm{s},x}^-\exp\{i(-k_{\mathrm{s},x}(z-z_{2N+1})-\omega t)\}
\end{eqnarray}
for the $x$ component and
\begin{eqnarray}
\label{Electricfield}
E_{y}=&A&_{\mathrm{s},y}^+\exp\{i(k_{\mathrm{s},y}(z-z_{2N+1})-\omega t)\} \nonumber \\
       &+&A_{\mathrm{s},y}^-\exp\{i(-k_{\mathrm{s},y}(z-z_{2N+1})-\omega t)\}
\end{eqnarray}
for the $y$ component, where
\begin{equation}
k_{\mathrm{s},x}=k_{\mathrm{s},y}=\frac{\omega}{c}n_\mathrm{s}.
\end{equation}
Using the characteristic matrix method \cite{BornWolf}, we have
\begin{equation}
\label{characteristicmatrix} \left(\begin{array}{c} A_{\mathrm{e},x}^+ \\
A_{\mathrm{e},x}^- \\ A_{\mathrm{e},y}^+ \\ A_{\mathrm{e},y}^-
\end{array}\right)=M
\left(\begin{array}{c} A_{\mathrm{s},x}^+ \\ A_{\mathrm{s},x}^- \\ A_{\mathrm{s},y}^+ \\
A_{\mathrm{s},y}^- \end{array}\right)
\end{equation}
where $M$ is a $4\times4$ matrix called the characteristic matrix of
the multilayer. This matrix can be calculated step by step by
solving numerically a $4\times4$ linear system of equations
corresponding to the appropriate boundary conditions that must be
fulfilled by the electric field at the interface between two
adjacent layers. Noting that
$A_{\mathrm{s},x}^-=A_{\mathrm{s},y}^-=0$ and taking
$A_{\mathrm{e},x}^+=1$ and $A_{\mathrm{e},y}^+=0$, we get
\begin{eqnarray}
\label{A_ex}
A_{\mathrm{e},x}^-&=&\frac{M_{21}}{\left(M_{11}-\frac{M_{13}M_{31}}{M_{33}}\right)}
- \frac{M_{23}M_{31}}{M_{33}\left(M_{11}-\frac{M_{13}M_{31}}{M_{33}}\right)} \\
\label{A_ey}
A_{\mathrm{e},y}^-&=&\frac{M_{41}}{\left(M_{11}-\frac{M_{13}M_{31}}{M_{33}}\right)}
-
\frac{M_{43}M_{31}}{M_{33}\left(M_{11}-\frac{M_{13}M_{31}}{M_{33}}\right)}.
\end{eqnarray}
The induced ellipticity per reflection $\psi_\mathrm{M}$ is then
given by
\begin{equation}
\label{alpha}
\tan{\psi_\mathrm{M}}=\frac{|A_{\mathrm{e},y}^-|}{|A_{\mathrm{e},x}^-|}.
\end{equation}

Since measured phase retardations presented in the previous section
are small, we only consider small birefringence. To fully reproduce
the experimental technique we calculate $\psi_\mathrm{M}$ as a
function of the angle between the polarization and the birefringent
axis of the simulated mirror. We checked that it behaves as a
standard wave plate from which we can extract the intrinsic phase
retardation $\delta_\mathrm{M}$.

\subsection{Results}
\label{CompRes}

Using the code based on the methods detailed in the previous
section, we have simulated several simple configurations. In the
trivial case in which every layer gives the same contribution to the
total effect, the straightforward result was that phase retardation
per reflection increases with the number of layers i.e. with the
mirror reflectivity. Random phase retardation and axis orientation
per layer has also been tested varying the range of variation of
these two parameters. No result similar to the experimental trend
has been obtained. The configurations which can reproduce this trend
are the ones in which the birefringent layers are only the ones
close to the substrate.

\begin{figure}
\begin{center}
\resizebox{1\columnwidth}{!}{
\includegraphics{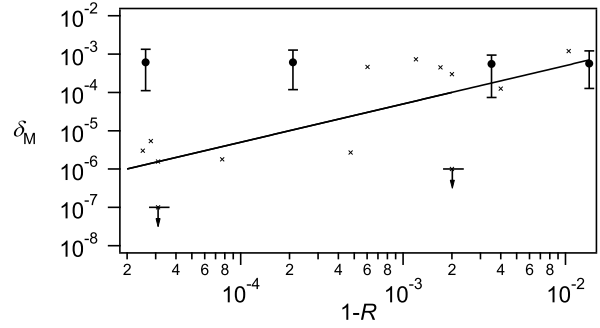}
} \caption{Two different numerical calculations for the induced
phase retardation per reflection as a function of $(1-R)$. Solid
curve : birefringence only for the first layer just after the
substrate. Dots with error bars : calculation with random
birefringence per each layer. Crosses : measurements plotted in
Fig.\,\ref{deltavsR}.} \label{Numerical_simulations}
\end{center}
\end{figure}

Figure \ref{Numerical_simulations} presents two different numerical
calculations for the induced phase retardation per reflection as a
function of $(1-R)$ where $R$ is the multilayer reflectivity we got
from our simulations. Crosses represent the measurements plotted in
Fig.\,\ref{deltavsR}. To match these experimental data, we have
chosen the parameters of our simulations such that numerical results
reproduce the experimental data for the highest $(1-R)$ available
value. Dots with error bars correspond to the result of random
calculations with $\delta n_\mathrm{L(H)}$ (resp. $\theta_j$)
randomly distributed inside the interval [$0,0.001$] (resp.
[$-\pi,\pi$]) for each layer. The error bar for each point
corresponds to the dispersion obtained with 10 tries. This result
does not reproduce the experimental data. On the other hand, the
solid curve has been obtained by including birefringence only for
the layer lying directly on the substrate. The parameters we used
are: $\delta n_\mathrm{H}=0.13$ for the $(2N+1)$th layer (zero for
the others). This result reproduces quite well the trend of the
experimental data i.e. the intrinsic phase retardation decreases by
3 orders of magnitude as $(1-R)$ decreases by 3 orders of magnitude.

\section{Conclusion}
\label{Conc}

Existing experimental data on interferential mirrors intrinsic phase
retardation, together with the two new measurements reported in this
work, clearly indicate that some physical effect decreases the
birefringence per reflection when the mirror reflectivity $R$
increases i.e. when one increases the number of layers used to
realize the interferential mirror. Our numerical calculations show
that it can be explained with a simple model in which only the
layers close to the substrate are birefringent. We could not find
any other reasonable configuration giving a trend similar to the
experimental one.

Our study cannot unveil the physical origin but it seems to indicate
in which part of the mirror the problem resides: the reflecting
layers  close to the substrate. We believe that it is a crucial
piece of information for mirror manufacturers in order to realize
birefringence free mirrors or at least to control and minimize the
effect.

Finally, although experimental data have been obtained by using
different mirrors that in principle have not been realized using
exactly the same manufacture protocol, we obtain a clear decreasing
of the phase retardation per refection as $R$ increases. But to
fully understand the origin of interferential mirror phase
retardation, we believe that next step should be to study a series
of mirrors, all made with the same industrial process, but with
different values of reflectivity $R$.

\section{Acknowledgements}
\label{Ack} This work has been performed in the framework of the BMV
project. We thank all the members of the BMV collaboration, and in
particular G. Bailly, T. Crouzil, J. Mauchain, J. Mougenot, G.
Tr\'enec. We acknowledge the support of the {\it ANR-Programme non
th\'{e}matique} (ANR-BLAN06-3-139634), and of the {\it
CNRS-Programme National Particule Univers}.

\end{document}